\documentclass{iaus}
\usepackage{graphicx}

\def\pa{\partial}
\def\vf{{\bf v}}
\def\Bf{{\bf B}}

\title[Origin of solar magnetism] 
{Origin of solar magnetism}

\author[Choudhuri]   
{Arnab Rai Choudhuri}

\affiliation{Department of Physics, Indian Institute of Science, Bangalore-560012\\ email: {\tt arnab@physics.iisc.ernet.in}} 

\pubyear{2010}
\volume{xxx}  
\pagerange{1--6}
\jname{Title of your IAU Symposium}
\editors{A.C. Editor, B.D. Editor \& C.E. Editor, eds.}
\begin{document}

\maketitle

\begin{abstract}
The most promising model for explaining the origin of solar magnetism
is the flux transport dynamo model, in which the toroidal field is
produced by differential rotation in the tachocline, the poloidal
field is produced by the Babcock--Leighton mechanism at the solar
surface and the meridional circulation plays a crucial role.  After
discussing how this model explains the regular periodic features of
the solar cycle, we come to the questions of what causes irregularities
of solar cycles and whether we can predict future cycles.  Only if
the diffusivity within the convection zone is sufficiently high,
the polar field at the sunspot minimum is correlated with strength
of the next cycle.  This is in conformity with the limited available
observational data.
\end{abstract}

\firstsection 
\section{Introduction}

The dynamo process taking place in the convection zone of the
Sun is believed to be the source of solar magnetism.
A solar dynamo model needs to explain two things.  Firstly,
it has to explain why the sunspot cycle is roughly periodic.
Secondly, it has to explain what causes the variabilities of
the cycles.  Why are not all the cycles identical?  We shall
address both these issues in this review. However, we would
like to emphasize that this is
{\it not} meant to be a 
comprehensive review of the solar dynamo. Limitation of space
forces us to concentrate only on certain very limited aspects of the solar
dynamo problem.  The readers are referred to the excellent reviews
by Ossendrijver (2003) and Charbonneau (2005) for discussions of
other aspects.

\section{Some basic considerations}

Polarities of sunspot pairs indicate a toroidal magnetic field
underneath the Sun's surface.  On the other hand, the magnetic field
of the Earth is of poloidal nature.  Parker (1955b) --- in perhaps
the most important paper on dynamo theory ever written ---
proposed that the sunspot cycle is produced by an oscillation
between toroidal and poloidal components, just as we see an
oscillation between kinetic and potential energies in a simple
harmonic oscillator.  This was a truly extraordinary suggestion
because almost nothing was known about the Sun's poloidal
field at that time.  Babcock \& Babcock (1955) were the first 
to detect the weak
\begin{figure}
\center
\includegraphics[width=12cm]{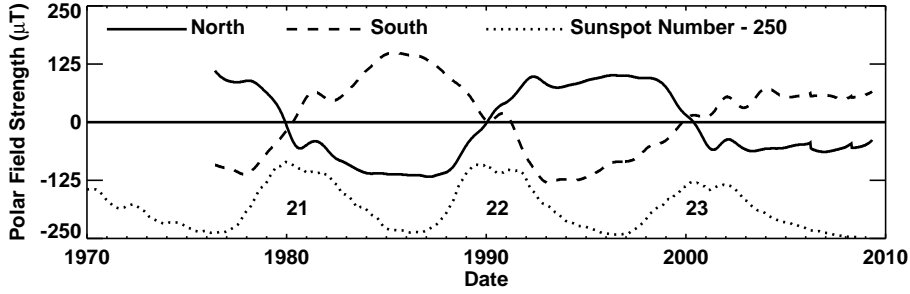}
\caption{The polar field of the Sun as a function of time
(on the basis of the Wilcox Solar Observatory data)
with the sunspot number shown below. From Hathaway (2010).}
\end{figure}
poloidal field having a strength of about 10 G near the Sun's
poles.  Only from mid-1970s we have systematic data of the
Sun's polar fields. Fig.~1 shows the polar fields of the Sun
plotted as a function of time, with the sunspot number plotted
below.  It is clear that the sunspot number, which is a 
proxy of the toroidal field, is maximum at a time when the polar
field is nearly zero.  On the other hand, the polar field is maximum
when the sunspot number is nearly zero.  This clearly shows
an oscillation between the toroidal and poloidal components,
as envisaged by Parker (1955b).  

\begin{figure}
\begin{minipage}[c]{0.55\textwidth}
\vspace{0.1cm}
\includegraphics[height=5.5cm,width=6cm]{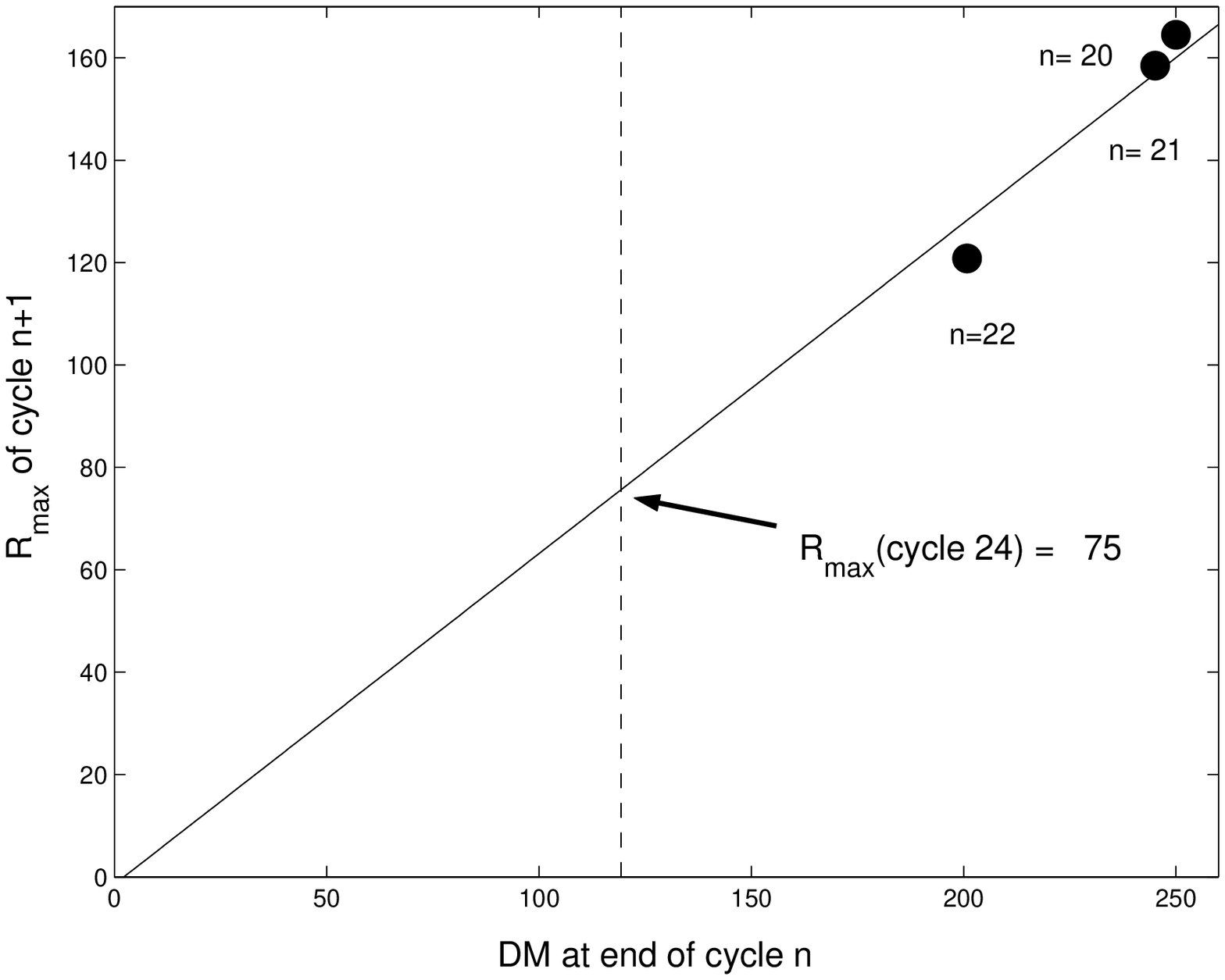}
\end{minipage}%
\hspace{0.3cm}
\begin{minipage}[c]{0.6\textwidth}
\includegraphics[height=5.5cm,width=6cm]{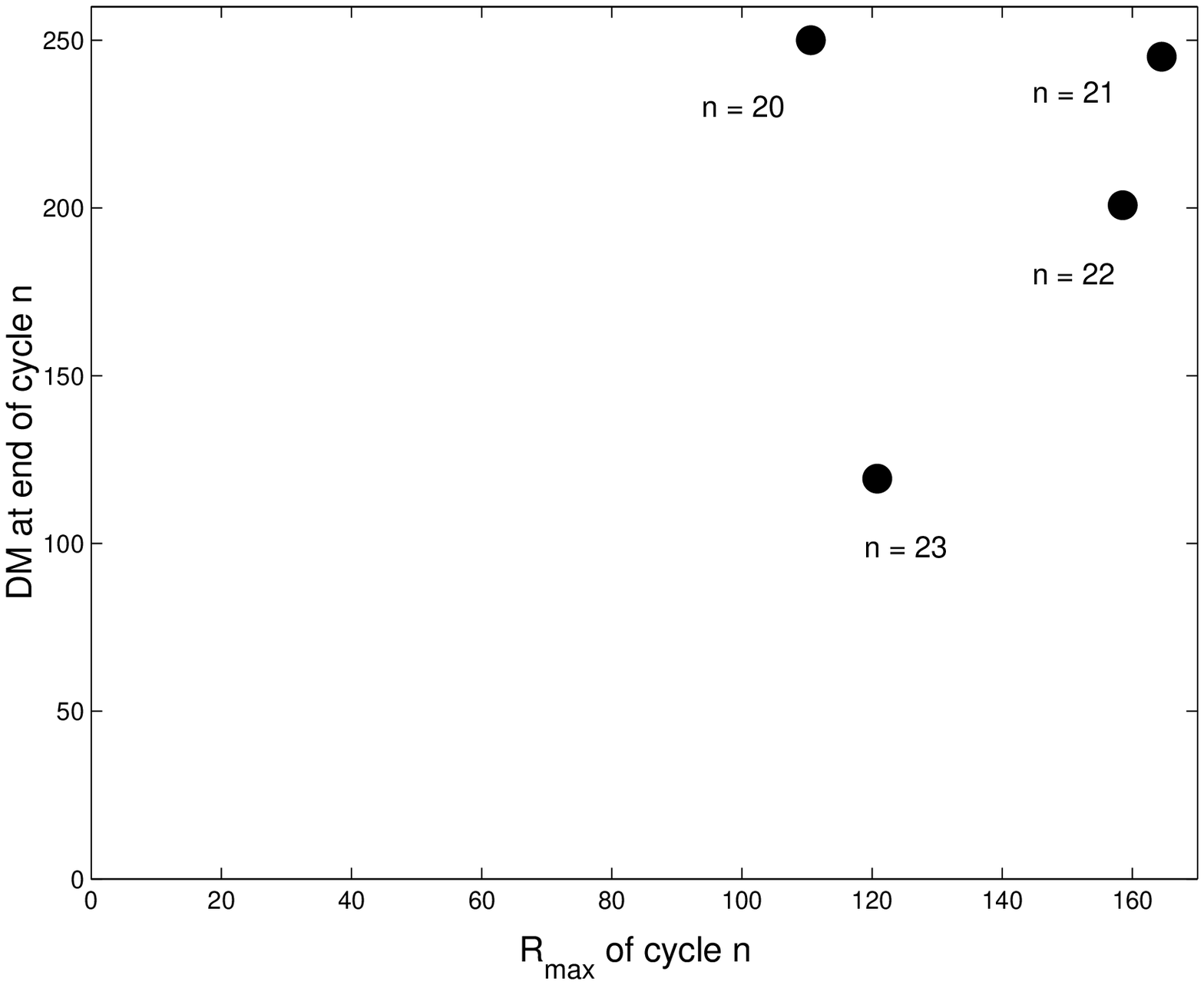}
\end{minipage}
\caption{The left panel shows a plot of the strength
of cycle $n+1$ against the polar field at the end of
cycle $n$. The right panel shows a plot of the polar field
at the end of cycle $n$ against the strength of the cycle
$n$. From Choudhuri (2008).}
\end{figure}

We note another important thing in Fig.~1.  The polar field at
the end of cycle~22 was weaker than the polar field in the previous
minimum.  We see that this weaker polar field was followed by
the cycle~23 which was weaker than the previous cycle.  Does this
mean that there is a correlation between the polar field during
a sunspot minimum and the next sunspot cycle?  In the left
panel of Fig.~2, we plot
the polar field in the minimum along the horizontal axis and
the strength of the next cycle along the vertical axis.  Although
there are only 3 data points so far, they lie so close to a
straight line that one is tempted to conclude that there is
a real correlation.  On the other hand, the right panel
of Fig.~2, which has
the cycle strength along the horizontal axis and the polar field
at the end of that cycle along the vertical axis, has points
which are scattered around.  Choudhuri, Chatterjee \& Jiang
(2007) proposed the following to explain these observations.
While an oscillation between toroidal and poloidal components
takes place, the system gets random kicks at the epochs indicated
in Fig.~3.  Then the poloidal field and the next toroidal field
should be correlated, as suggested by the left panel of
Fig.~2.  On the other
hand, the random kick ensures that the toroidal field is not
correlated with the poloidal field coming after it, as seen
in the right panel of Fig.~2.

If there is really a correlation between the polar field at the
minimum and the next cycle, then one can use the polar field
to predict the strength of the next cycle (Schatten et al.\
1978).  Since the polar field in the just concluded minimum
has been rather weak, several authors (Svalgaard, Cliver \& Kamide 2005;
Schatten 2005) suggested that the coming cycle~24 will be rather
weak.  Very surprisingly, the first theoretical prediction based
on a dynamo model made by Dikpati \& Gilman (2006) is that the
cycle~24 will be very strong.    Dikpati \& Gilman (2006)
assumed the generation of the poloidal field from the toroidal
field to be deterministic, which is not supported by observational
data shown in the right panel
of Fig.~2. Tobias, Hughes \& Weiss (2006) make the
following comment on this work:
``Any predictions made with such models should be treated with extreme 
caution (or perhaps disregarded), as they lack solid physical underpinnings.''
While we also consider many aspects of the Dikpati--Gilman work wrong which
will become apparent to the reader, we 
cannot also accept the opposite extreme viewpoint
of Tobias, Hughes \& Weiss (2006), who suggest that the solar dynamo is a
nonlinear chaotic system and predictions are impossible or useless. If that
were the case, then we are left with no explanation for the correlation
seen in the left panel of Fig.~2.  

\begin{figure}
\center
\includegraphics[width=12cm]{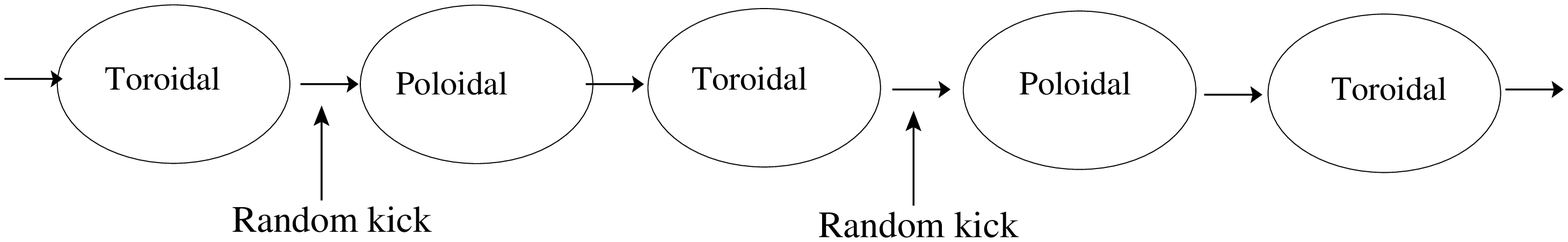}
\caption{A schematic cartoon of the oscillation between toroidal
and poloidal components, indicating the epochs when the system
is subjected to random kicks.}
\end{figure}

\section{The flux transport dynamo model of the solar cycle}

We now come to the theoretical question as to what causes the
oscillation between the toroidal and poloidal components.  Then
in \S4 we shall discuss how the random kicks shown in Fig.~3 arise.

The differential rotation of the Sun stretches
any poloidal field line to produce a toroidal field, primarily
in the tachocline where the differential rotation is
strongest.  Then this
toroidal field rises due to magnetic buoyancy and produces
sunspots (Parker 1955a).  It is fairly easy to see how the toroidal field
can be produced from the poloidal field.  The production of
the poloidal field from the toroidal field is more non-trivial.
The original idea of Parker (1955b) and Steenbeck, Krause \&
R\"adler (1966) was that the toroidal field is twisted by the
helical turbulence in the convection zone giving rise to a
field in the poloidal plane.  This is, however, possible only
if the toroidal field is not stronger than its equipartition
value.  Simulations based on the thin flux tube equation (Spruit
1981; Choudhuri 1990) suggest that the toroidal field is likely
to be as strong as $10^5$ G at the bottom of the convection
zone (Choudhuri 1989; D'Silva \& Choudhuri 1993)
and cannot be twisted by helical turbulence. We therefore
invoke the alternate idea proposed by Babcock (1961) and Leighton (1969).
We know that bipolar sunspots on the solar surface have tilts
with respect to latitude lines, the leading sunspot appearing
nearer the equator. Suppose we consider a pair in which
the positive polarity sunspot is nearer the equator.  When
this pair decays, more positive polarity is spread around in
lower latitudes and more negative polarity in the higher latitudes.
This causes a poloidal magnetic field.  A sunspot pair which
forms from the toroidal field, therefore, acts as a conduit
through which a conversion from the toroidal field to the 
poloidal field takes place.  In a nutshell,  
the differential rotation acting on the poloidal field
produces the toroidal field.  The toroidal field, in turn,
gives rise to the poloidal field by the Babcock--Leighton 
mechanism, thereby completing the cycle.  However, a theoretical
dynamo model with differential rotation and Babcock--Leighton
mechanism alone gives rise to belts of toroidal field
propagating poleward, in contradiction to the observation
that the sunspots belts propagate equatorward.  So we need
something else.  Choudhuri, Sch\"ussler \& Dikpati (1995),
who pointed out this difficulty, also proposed a solution.  They
realized that the meridional circulation of the Sun can turn
things around and can make the dynamo wave propagate in the
correct direction.

\begin{figure}
\center
\includegraphics[width=7cm]{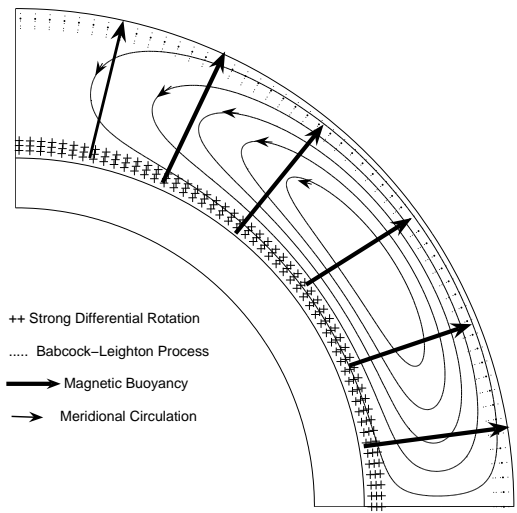}
  \caption{A cartoon explaining how the flux transport dynamo works.}
\end{figure}

Fig.~4 schematically shows the dynamo in which the meridional
circulation plays a key role.  Such a dynamo is often called a
{\em flux transport dynamo}.  First two-dimensional models of
the flux transport dynamo were constructed by
Choudhuri, Sch\"ussler \& Dikpati (1995) and Durney (1995),
although the basic ideas were given in an early paper by
Wang, Sheeley \& Nash (1991). 
We now write down the basic equations of the flux transport dynamo.
The axisymmetric magnetic field is represented in the form
$$
{\bf B} = B (r, \theta,t) {\bf e}_{\phi} + \nabla \times [ A
(r, \theta,t) {\bf e}_{\phi}],
\eqno(1)$$
where $B (r, \theta,t)$ and $A (r, \theta,t)$ respectively
correspond to the toroidal and poloidal components.  They
evolve according to the equations
$$\frac{\pa A}{\pa t} + \frac{1}{s}(\vf.\nabla)(s A)
= \eta_{p} \left( \nabla^2 - \frac{1}{s^2} \right) A + \alpha B,
\eqno(2)$$
$$ \frac{\pa B}{\pa t} 
+ \frac{1}{r} \left[ \frac{\pa}{\pa r}
(r v_r B) + \frac{\pa}{\pa \theta}(v_{\theta} B) \right]
= \eta_{t} \left( \nabla^2 - \frac{1}{s^2} \right) B 
+ s(\Bf_p.\nabla)\Omega + \frac{1}{r}\frac{d\eta_t}{dr}
\frac{\partial}{\partial{r}}(r B), \eqno(3)$$
where $s = r \sin \theta$ and the other symbols have the
usual meanings.
Since nothing much can be done analytically, these equations
have to be solved numerically.  We have developed a code {\em 
Surya}, which we make available to anybody upon request. 
The right panel of Fig.~5, taken from Chatterjee, Nandy \& Choudhuri (2004),
shows the theoretical butterfly diagram superposed on a time-latitude
plot indicating how $B_r$ evolves at the solar surface.
The equatorward propagating meridional circulation at the
bottom of the convection zone forces the sunspot belts
to migrate equatorward, whereas the poleward propagating
meridional circulation at the surface advects the poloidal
field poleward.  The right panel of 
Fig.~5 has to be compared with the comparable
plot of observational data shown in the left panel.  Given the fact that
this was one of the first efforts of matching observational
data in such detail, the fit is not too bad.

\begin{figure}
\begin{minipage}[c]{0.55\textwidth}
\vspace{0.1cm}
\includegraphics[height=6.5cm,width=9.5cm]{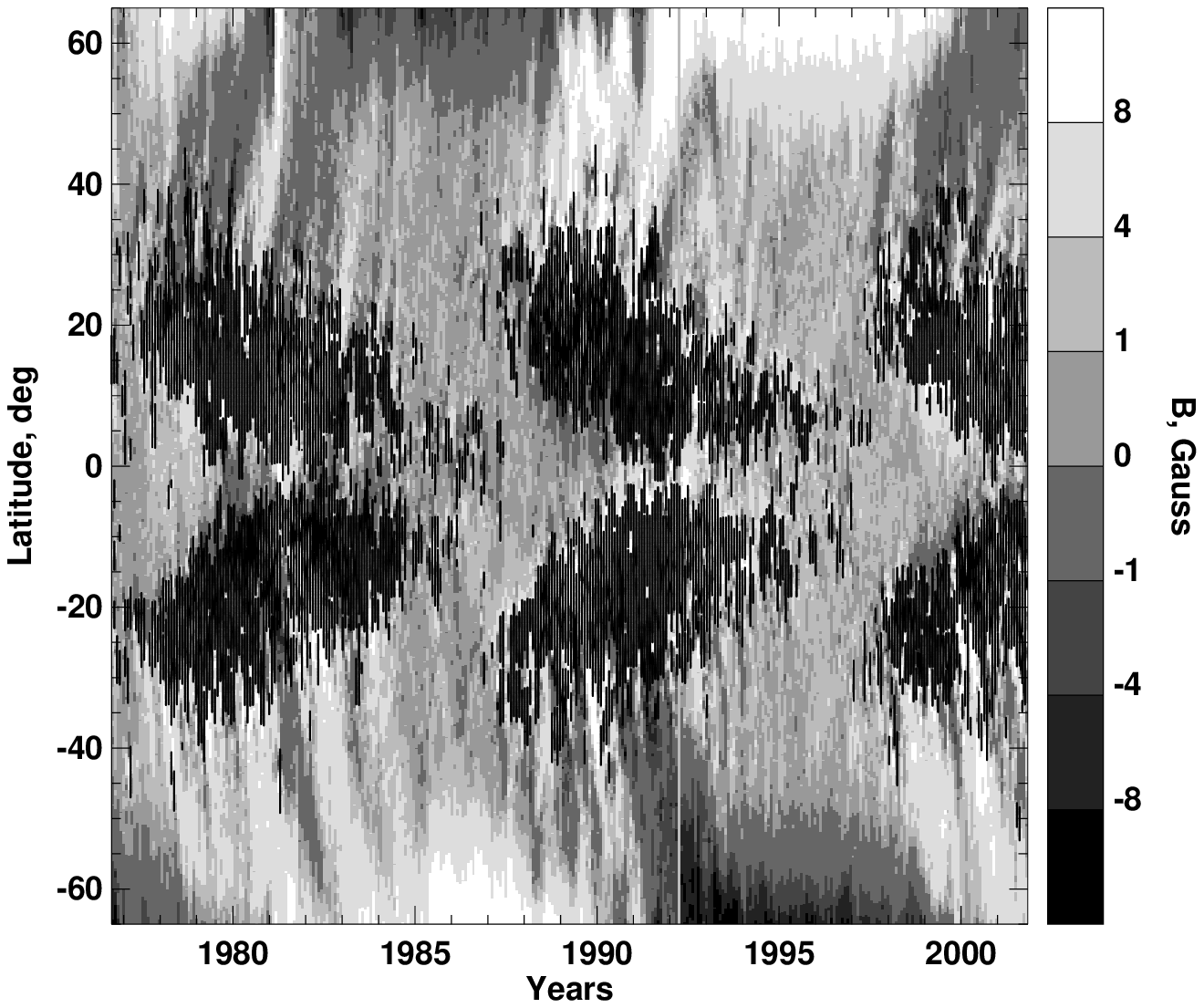}
\end{minipage}%
\hspace{0.3cm}
\begin{minipage}[c]{0.6\textwidth}
\includegraphics[height=6cm,width=6.0cm]{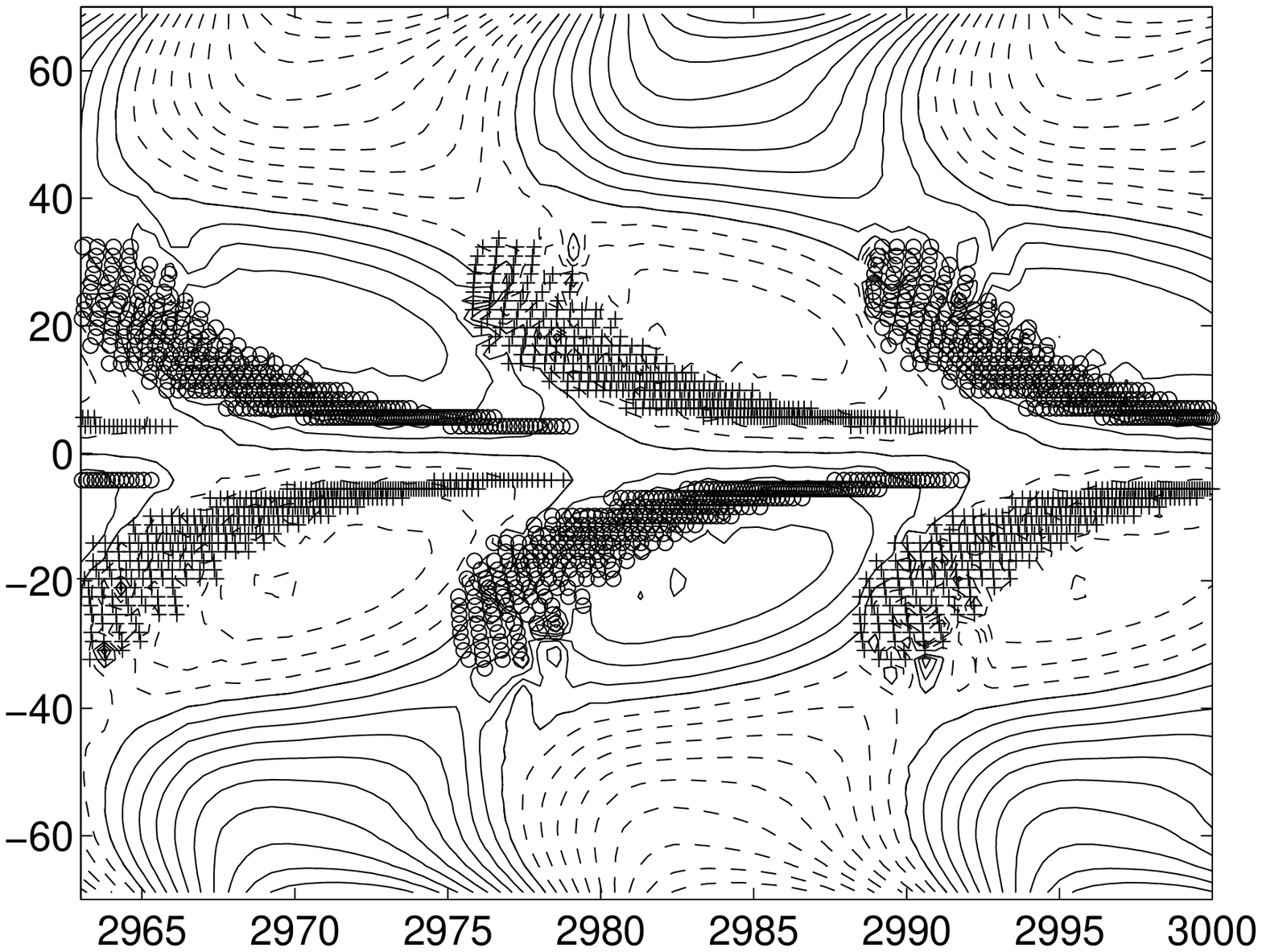}
\end{minipage}
\caption{Butterfly diagram of sunspots superposed on the time-latitude
  plot of $B_r$. The observational plot is shown on the left.  
  The comparable theoretical
  plot obtained by the dynamo model of Chatterjee, Nandy \& Choudhuri 
(2004) is on the right.}
\end{figure}

The original flux transport dynamo model of Choudhuri, Sch\"ussler
\& Dikpati (1995) led to two offsprings: a high diffusivity model
and a low diffusivity model.  The diffusion times in these two models
are of the order of 5 years and 200 years respectively.  The high
diffusivity model has been developed by a group working in IISc Bangalore
(Choudhuri, Nandy, Chatterjee, Jiang, Karak), whereas the low diffusivity
model has been developed by a group working in HAO Boulder (Dikpati,
Charbonneau, Gilman, de Toma).  The differences between these models
have been systematically studied by Jiang, Chatterjee \& Choudhuri
(2007) and Yeates, Nandy \& Mckay (2008).  Both these models are
capable of giving rise to oscillatory solutions resembling solar
cycles.  However, when we try to study the variabilities of the
cycles, the two models give completely different results.  We need
to introduce fluctuations to cause variabilities in the cycles.
In the high diffusivity model, fluctuations spread all over the
convection zone in about 5 years.  On the other hand, in the low
diffusivity model, fluctuations essentially remain frozen during
the cycle period.  Thus the behaviours of the two models are totally
different on introducing fluctuations.  

\section{Modelling irregularities of solar cycles}

Let us now finally come to question as to what produces the variabilities
of cycles and whether we can predict the strength of a cycle before its advent.
Some processes in nature can be predicted and some not.  We can easily
calculate the trajectory of a projectile by using elementary mechanics.
On the other hand, when a dice is thrown, we cannot predict which side
of the dice will face upward when it falls.  Is 
the solar dynamo more like the trajectory 
of a projectile or more like the throw of a dice? Our point of view is that the
solar dynamo is not a simple unified process, but a complex combination of
several processes, some of which are predictable and others not.  Let
us look at the processes which make up the solar dynamo.
 
The flux transport dynamo model combines three basic processes. (i) The
strong toroidal field is produced by the stretching of the poloidal
field by differential rotation in the tachocline. 
(ii) The toroidal field generated in the tachocline
gives rise to active regions due to magnetic buoyancy and then the decay
of tilted bipolar active regions produces the poloidal field by the 
Babcock--Leighton mechanism.  (iii) The poloidal field is
advected by the meridional circulation first to high latitudes and then down
to the tachocline, while diffusing as well.   We believe that
the processes (i) and (iii) are reasonably ordered and deterministic.
In contrast, the process (ii) involves an element of randomness due
to the following reason.  The poloidal field produced from the decay of 
a tilted bipolar region by the Babcock--Leighton process depends on
the tilt.  While the average tilt of bipolar regions at a certain
latitude is given by Joy's law, we observationally find quite a large
scatter around this average.
Presumably  the action
of the Coriolis force on the rising flux tubes gives rise to Joy's
law (D'Silva \& Choudhuri 1993), whereas convective buffeting of 
the rising flux tubes in the upper layers of the convection zone causes the
scatter of the tilt angles (Longcope \& Choudhuri 2002). 
This scatter in the tilt angles certainly introduces a
randomness in generation process of the poloidal field
from the toroidal field.  Choudhuri,
Chatterjee \& Jiang (2007) identified it as the main source of
irregularity in the dynamo process, which is in agreement 
with Fig.~3. It may be noted that Choudhuri (1992) was the
first to suggest that the randomness in the poloidal field
generation process is the source of fluctuations in the dynamo.

\begin{figure}
\center
\includegraphics[width=4cm]{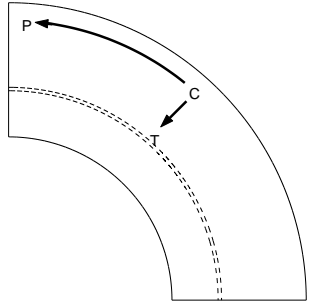}
\caption{A sketch
indicating how the poloidal field produced at  C during a
maximum gives rise to the polar field at P during the
following minimum and the toroidal field at T during the next
maximum. From Jiang, Chatterjee \& Choudhuri (2007).}
\end{figure}

The poloidal field gets built up during the declining phase of
the cycle and becomes concentrated near the poles during the
minimum.
The polar field at the solar minimum produced in a theoretical 
mean field dynamo
model is some kind of `average' polar field during a typical
solar minimum.  The observed polar field during
a particular solar minimum may be stronger or weaker than this average 
field. The theoretical dynamo model has to be updated by feeding
the information of the observed polar field in an appropriate
way, in order to model particular cycles.
Choudhuri, Chatterjee \& Jiang (2007) proposed to model this
in the following way. They ran the dynamo code from a
minimum to the next minimum in the usual way.  After stopping
the code at the minimum,
the poloidal field of the theoretical model was multiplied
by a constant factor everywhere above $0.8 R_{\odot}$ to bring
it in agreement with the observed poloidal field.  Since
some of the poloidal field at the bottom of the convection zone may
have been produced in the still earlier cycles, it is left unchanged
by not doing any updating below $0.8 R_{\odot}$.  Only the poloidal
field produced in the last cycle which is concentrated in the upper
layers gets updated. After 
this updating, we run the code till the next minimum, when the code
is again stopped and the same procedure is repeated. 
Our solutions are now no longer
self-generated solutions from a theoretical model alone, but are
solutions in which the random aspect of the dynamo process has been
corrected by feeding the observational data of polar fields into
the theoretical model.

Before presenting the results obtained with this procedure, we come
to the question how the correlation between the polar field at the
minimum and the strength of the next cycle as seen in the left panel of Fig.~2 may
arise.  This was first explained by Jiang, Chatterjee \& Choudhuri
(2007). The Babcock--Leighton process would first produce the poloidal
field around the region C in Fig.~6.  Then this poloidal field will
be advected to the polar region P by meridional circulation and will
also diffuse to the tachocline T. In the high diffusivity model, this
diffusion will take only about 5 years and the toroidal field of the
next cycle will be produced from the poloidal field that has diffused
to T.  If the poloidal field produced at C is strong, then both the
polar field at P at the end of the cycle and the toroidal field at T
for the next cycle will be strong (and vice versa).  We thus see that
the polar field at the end of a cycle and the strength of the next
cycle will be correlated in the high diffusivity model.  But this
will not happen in the low diffusivity model where it will take more
than 100 years for the poloidal field to diffuse from C to T
and the poloidal field will reach the tachocline only due to the
advection by meridional circulation taking a time of about 20 years. If we
believe that the 3 data points in the left panel of 
Fig.~2 indicate a real correlation,
then we have to accept the high diffusivity model!

\begin{figure}
\center
\includegraphics[width=11cm,height=5.5cm]{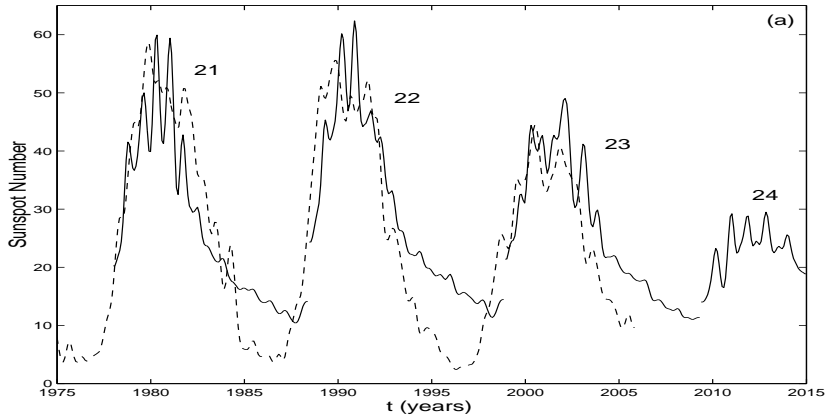}
\caption{The theoretical monthly sunspot number 
(solid line) for the last few years as well as 
the upcoming next cycle, plotted along with the observational data (dashed line)
for the last few years. From Choudhuri, Chatterjee and
Jiang (2007).}
\end{figure}

Finally the solid line in
Fig.~7 shows the sunspot number calculated from our high
diffusivity model (Choudhuri, Chatterjee \& Jiang 2007). Since
systematic polar field measurements are available only from the
mid-1970s, the procedure outlined above could be applied only from
that time.  It is seen from Fig.~7 that our model matches the last
three cycles (dashed line) reasonably well and predicts a weak cycle 24.  

\section{Conclusion}

It seems that the flux transport dynamo model explains many aspects
of solar magnetism and probably will stand the test of time as the
appropriate theoretical model for the solar cycle.  In the recent
past, two versions of the flux transport dynamo model have been
developed in considerable detail: the high diffusivity version and
the low diffusivity version.  In the high diffusivity version, the
poloidal field created at the surface by the Babcock--Leighton process
is transported to the tachocline by diffusion.  On the other hand,
this transport is affected by the meridional circulation in the
low diffusivity version of the flux transport dynamo.  Several authors
(Chatterjee, Nandy \& Choudhuri 2004; Chatterjee \& Choudhuri 2006;
Jiang, Chatterjee \& Choudhuri 2007; Goel \& Choudhuri 2009;
Hotta \& Yokoyama 2010a, 2010b; Karak
\& Choudhuri 2010) have given several independent arguments in support
of the high diffusivity model. This model has also been applied to
provide theoretical explanations of such things as the helicity of
active regions (Choudhuri, Chatterjee \& Nandy 2004), the torsional
oscillations (Chakraborty, Choudhuri \& Chatterjee 2009) and the
Maunder minimum (Choudhuri \& Karak 2009).  However, a crucial test
of the high diffusivity model lies ahead.     
Since the high diffusivity model makes the polar field at a minimum
correlated with the strength of the next cycle and the polar field
during the immediate past minimum was weak, an inescapable conclusion
of the high diffusivity model is that the upcoming cycle will be weak
(Choudhuri, Chatterjee \& Jiang 2007; Jiang, Chatterjee \& Choudhuri 2007).
While the early indications are that this will indeed be the case, we
need to wait for a couple of years to be absolutely sure.  If cycle~24
turns out to be strong as predicted by Dikpati \& Gilman (2006), 
then there will be no way of reconciling that fact with the high
diffusivity model and the model will have to be discarded.  On the
other hand, if cycle~24 turns out to be weak, that will provide a
further corroboration of the high diffusivity model.

\end{document}